\documentclass[11pt,a4paper]{article}
\begin{document}

\title{New Investigation on the Spheroidal
Wave Equations  \footnote{ E-mail: tgh-2000@263.net,
tgh20080827@gmail.com}
\author{Guihua Tian
\\
School of Science, Beijing University of Posts And
Telecommunications.
\\ Beijing 100876 China.}}
\date{April 9, 2010}
\maketitle
\begin{abstract}
Changing the spheroidal wave equations into new
Schr$\ddot{o}$dinger's form, the super-potential expanded in the
series form of the parameter $\alpha$are obtained in the paper. This
general form of the super-potential makes it easy to get the ground
eigenfunctions of the spheroidal equations. But the shape-invariance
property is not retained and the corresponding recurrence relations
of the form (\ref{p-n m fixed}) could not be extended from the
associated Legendre functions to the case of the spheroidal
functions.

\textbf{PACs: 03.65Ge, 02.30Gp,11.30Pb}
\end{abstract}

\section*{Introduction}

  \ \ \ \ Spheroidal wave functions appear in many different context
  in physics and mathematics \cite{flammer}-\cite{slepian}. Their differential
  equations are \begin{equation} \left[\frac{d}{dx}\left[(1-x^2) \frac{d}{d
x}\right]+E+ \alpha x^2
 -\frac{m^{2}}{1-x^2} \right]\Theta=0, x\in (-1,+1).\label{3}
\end{equation}
They just have one more term $\alpha x^2$ than the spherical wave
equations (the associated Legendre' equations). This extra term
makes great difference in Equations: the spherical wave equations
belong to the case of the confluent super-geometrical equations with
one regular and one irregular singularities, whereas the spheroidal
wave equations are the confluent Heun equations containing two
regular and one irregular singularities. The extra singularity makes
them one of the toughest problems for researchers to
treat\cite{flammer}-\cite{li}.

The spheroidal functions are the solutions  of the Sturm-Liouville
eigenvalue problem of Equation (\ref{3}) with the natual condition
$\Theta $ finite at the boundaries $x=\pm 1$. The eigenvalues are
the allowable values $E_0,\ E_1,\dots, E_n,\dots$ of the parameter
$E $;  and the spheroidal wave functions are the corresponding
eigenfunctions $\Theta_0,\Theta_1,\dots,
\Theta_n,\dots$\cite{flammer}-\cite{li}.

Though the spheroidal functions are the extension of the associated
Legendre-functions $P_l^m(x)$, they stand in vivid contrast against
the associated Legendre-functions. The associated Legendre-functions
$P_l^m(x)$ have mane elegant properties:
\begin{enumerate}
  \item{When $m=0$, the Legendre-functions $P_n(x)=P_n^0(x)$ are polynomials of the independent variable $x$;}
  \item{Back to the variable $\theta $ with $x=\cos\theta $, the recursion relation among the Legendre-functions could be written as:
  \begin{eqnarray} P_{n}(\cos\theta )=n^{-1} \left[-n\cos\theta -\sin\theta \frac{d}{d\theta }\right]P_{n-1}(\cos\theta ),\
  n=1,2,3,\dots,\end{eqnarray}
  so, all $P_n(\cos\theta )$ could be deduced from the ground
function $P_0$ from the recursion relations }
  \item{With the quantity $m$ fixed, all the associated
Legendre-functions $P_n^m(\cos\theta )$ could be derived from
$P_m^m$ by
\begin{eqnarray}
P_n^m(\cos\theta )&=& \left[(n+m)(n-m)\right]^{-\frac12}\left[-n\cos\theta -\sin\theta \frac{d}{d\theta }\right]P_{n-1}^m(\cos\theta ),\nonumber\\
  n&=&m+1,m+2,m+3,\dots,\label{p-l}.\end{eqnarray} }

  \item{With the quantity $n$ fixed and the transformation $P_n^m=\frac{\Psi_n(\theta,m)}{\sin^{\frac12}
\theta}$, the recurrence relations between adjoint $m$ of the
associated Legendre-functions become
  \begin{eqnarray} \Psi_n(\theta, m+1) &=& \left[(n+m+1)(n-m)\right]^{-\frac12} *
  \left[(m+\frac12)\cot\theta -\frac{d}{d\theta }\right]\Psi_{n}(\theta,m ),\nonumber\\
  m&=&0,1,2,\dots, n\label{p-n m fixed}.\end{eqnarray}} These equations (\ref{p-n m
  fixed}) are more useful than that of (\ref{p-l}), they result in
  the relation \begin{eqnarray} P_n^m(x)=\left(1-x^2\right)^{\frac
m2}\frac{d^mP_n(x)}{dx^m}\label{p-m}.  \end{eqnarray}
  \end{enumerate}
The associated Legendre-functions $P_n^m(x)$ are related to the
Legendre's functions by (\ref{p-m}), which is the repeated forms of
the relations (\ref{p-n m fixed}). The significant meaning of these
relations are that one only needs to know the Legendre's functions
to obtain the associated Legendre's functions. So the two important
works of the associated Legendre's equations lie in finding the
ground state function of the Legendre's equations, that is, the case
of $m=0$ and the recurrence relations of the type of Eqs.(\ref{p-l})
and (\ref{p-n m fixed}). Eqs.(\ref{p-l}) could be used to get the
excited state eigenfunctions of the Legendre's equations; while
Eqs.(\ref{p-n m fixed}) make it possible to obtain the associated
Legendre's functions from the Legendre's functions.

Like the spherical wave functions, the spheroidal wave functions are
important functions
  applied extensively in many different branches
  in physics and mathematics \cite{flammer}-\cite{slepian}, hence, it is natural for one to require
  whether or not the similar
relations exist for the spheroidal functions. This is a long
standing problem. However, it has actually been treated in the
series papers\cite{tian first}-\cite{tang}. In these papers, the
supersymmetric quantum mechanics (SUSYQM) is first applied to study
them, and new interesting results are obtained\cite{tian}-\cite{tian
general}: (1)the general form of the the ground eigenvalue and
eigenfunction are given, which reduce to the ground eigenvalue
$m(m+1)$ and the ground eigenfunction $P_m^m$ when $\alpha=0$; (2)
the generalized recurrence relations like that of (\ref{p-l}) are
obtained for the spheroidal functions \cite{tian general}. See
references \cite{tian}-\cite{tian general} for details.
 As stated before, the relations (\ref{p-n m fixed}) are crucial for one to obtain
 the associated Legendre's functions from the Legendre's
 polynomials. Therefore, it is natural for one to investigate whether the
 same kind recurrence relations could be extended to the case of the spheroidal
 functions. This is just what the present paper tries to do.

As done before, the supersymmetry quantum mechanics is used to deal
with the problem. The recurrence relations turn out to be the
shape-invariance relations for the corresponding differential
equation. Thus, the spheroidal equations are first transformed into
the Schr$\ddot{o}$dinger equations. Then the super-potential $W$ is
introduced and solved in the series form of the parameter $\alpha$.
Thirdly, the ground state eigenfunction is obtained. Finally, the
shape-invariance property is checked, and the recurrence relations
are studied. In the following, the same steps will proceed too.

  In use of the perturbation methods in super-symmetry quantum mechanics,
  it is necessary to  rewrite the differential equations in the
Schr$\ddot{o}$dinger form. In the previous papers,  the original
form \begin{equation} \left[\frac{1}{\sin
\theta}\frac{d}{d\theta}\left(\sin \theta \frac{d}{d \theta}\right)+
\alpha\cos ^2 \theta -\frac{m^{2}}{\sin ^2
\theta}\right]\Theta=-E\Theta\label{2}
\end{equation} is used to study. It is obtained from Eq.(\ref{3}) by
the transformation $x=\cos\theta$ and the corresponding boundary
conditions become $\Theta$ finite at $\theta=0,\ \pi$.

There are two ways to transform Eq.(\ref{2}) or (\ref{3}) into the
forms of the Schr$\ddot{o}$dinger equation. The first transformation
is to change the eigenfunction through the transformation $\Theta
=\frac{\Psi}{\sin^{\frac12} \theta}$, and the Schr$\ddot{o}$dinger's
form becomes
\begin{eqnarray} \frac{d^2\Psi}{d\theta^2}+\left[\frac14+ \alpha\cos
^2 \theta
 -\frac{m^{2}-\frac14}{\sin ^2
\theta}+E \right]\Psi=0\label{main Eq}.
\end{eqnarray}
 The corresponding boundary conditions now are
$\Psi(0)=\Psi(\pi)=0$.

  In the series papers \cite{tian}-\cite{tian general}, the supersymmetric quantum mechanics
(SUSYQM) is first applied to treat Eqs.(\ref{main Eq}). The focus is
the super-potential $W$, which is determined by the potential
\begin{eqnarray} V(\theta,\alpha,m)=-
 \left[\frac14+ \alpha\cos
^2 \theta-\frac{m^{2}-\frac14}{\sin ^2 \theta}
\right]=W^2-\frac{dW}{d\theta}.\end{eqnarray} The ground function is
connected with the super-potential $W$ by
\begin{equation} \Psi_0=N\exp\left[-\int
Wd\theta\right].\end{equation} Hence, whenever the super-potential
is known, the ground state function is known too
\cite{tian}-\cite{tian general}. By the perturbation methods in the
super-symmetry quantum mechanics, it is the super-potential $W$ that
is expanded in the series form of the parameter $\alpha$, that is,
$W=\sum_{n=0}^{\infty}\alpha^nW_n$. By this method, new interesting
results are obtained\cite{tian}-\cite{tian general}: the first
several terms of the the ground eigenvalue and eigenfunction are
given, which reduce to the ground eigenvalue $m(m+1)$ and the ground
eigenfunction $P_m^m$ of the associated Lengdre' functions. Of
course, these results are obtained through the corresponding terms
for the super-potential $W$\cite{tian}-\cite{tian
first}:\begin{eqnarray} &&W_0=-\left(m+\frac12\right) \cot\theta,\
E_{00}=m(m+1).\label{w-0 t}\\&&
W_1=\frac{\sin\theta\cos\theta}{2m+3}\label{good result of w1
t}\\&&W_2= \left[\frac{-\sin\theta
\cos\theta}{(2m+3)^3(2m+5)}+\frac{\sin^3\theta
\cos\theta}{(2m+3)^2(2m+5)}\right] \label{w2 in complex
t}.\end{eqnarray}
 Later the reference \cite{tang} generalized
the results and obtained the general form for the super-potential
$W=\sum_{n=0}^{\infty} W_n\alpha^n$ as \begin{eqnarray}
W_n=\sum_{k=1}^{n}\frac{\hat{a}_{n,k}}{(2m+3)^n}\sin\theta\cos^{2k+1}\theta,
n=1,2,\ldots \label{tang's w}\end{eqnarray} where $\hat{a}_{n,k}$
could be easily determined step by step. The shape-invariance
property is also proved in it. The form of the super-potential in
(\ref{tang's w}) is a little different in form from that of the
references \cite{tian}. The reference \cite{tian general} directly
generalized the form of the reference \cite{tian} and gived an
alternative form of the super-potential as
\begin{eqnarray}W_n=\cos\theta
\sum_{k=1}^{n}\tilde{a}_{n,k}\sin^{2k-1}\theta. \label{wn to be
proved}\end{eqnarray}

Our motivation is to investigate whether we could extend the
relations of the type (\ref{p-n m fixed}) to the cases of the
spheroidal functions. In these recurrence relations, it is the
integer $m$ that is different. From the point of the super-symmetry
quantum mechanics, the recurrence relations are oriented in the
properties of the shape-invariance of the super-potential and are
the relations between different eigen-functions of correspondingly
different eigen-values. Therefore, the integer $m$ should stand in
the position of the eigen-values, the energies. Actually, this could
not be met in Eq.(\ref{2}) where it is the quantity $E$ that stands
as the eigen-value. However, there is another way to transform
Eq.(\ref{2}) to the Schr$\ddot{o}$dinger form.  Changing the
independent variable $\theta$ to the new one
$z=\lg\tan(\frac{\theta}2)$, the new Schr$\ddot{o}$dinger's form for
the spheroidal wave equations is obtained as:
\begin{eqnarray}
\frac{d^2\Theta}{dz^2}+\left[E{\rm sech}^2z+ \alpha{\rm
sech}^2z-\alpha{\rm sech}^4z-m^2 \right]\Theta=0.\label{main Eq2}
\end{eqnarray}
First, one notices that the interval $(0,\pi)$ in the original
variable $\theta$ now corresponds to the interval $(-\infty,
+\infty)$ in the new variable $z$. Secondly, the boundary conditions
now turn out as $\Theta$ finite at $z\rightarrow \pm \infty$. The
most important thing is that Eq (\ref{main Eq2}) makes the term
containing the original eigenvalue $E$ no longer in the position of
the eigenvalue. Instead, it is the quantity $-m^2$ now that is the
eigenvalue of Eq.(\ref{main Eq2}). This is just what we want to
obtain. It will beneficial for one to further investigate the kind
relations \ref{p-n m fixed} for the spheroidal functions.

In the above Eq.(\ref{main Eq2}), the potential is
\begin{eqnarray} V(z,\alpha)=-\left[E{\rm sech}^2z+
\alpha{\rm sech}^2z-\alpha{\rm sech}^4z\right].\label{potential in
z}\end{eqnarray}

When $\alpha=0$, Eq.(\ref{main Eq2}) is just the form from the
associated Legendre equations (\ref{2}); its ground energy is
$-m^2$, and the nodeless ground eigenfunction is the associated
Legendre function $P_m^m$ with the original eigenvalue $E$ taking
$m(m+1)$. Therefore, when $\alpha\ne 0$, the ground energy is also
$-m^2$ and the ground eigenfunction is $\Theta(z,m,E)$, which must
be nodeless. Actually the original eigenvalue $E$ could be
determined in the following by requiring the eigenfunction finite at
the infinities. Its value could also be obtained from the previous
papers \cite{tang}, \cite{tian general}.

The super-potential $W$ is determined from the potential
$V(z,\alpha)$ by \begin{equation}
W^2-\frac{dW}{dz}=V(z,\alpha)+m^2\label{potential and w relation in
z}
\end{equation}
where subtracting the ground eigenvalue $-m^2$ to make the potential
factorable. This  equation is the same difficult to treat as the
Schr$\ddot{o}$dinger original Eq.(\ref{main Eq2}), hence, the
perturbation method is used to solve it. First, the super-potential
$W$ is expanded as the series of the parameter $\alpha$, that is,
\begin{equation}
W=W_0+\alpha W_1+\alpha ^2 W_2+\alpha ^3 W_3+\ldots
=\sum_{n=0}^{\infty}\alpha^nW_n.\label{W}
\end{equation}
\begin{eqnarray}
 W^2&-&W'= W_0^2-W'_0+\alpha \left(2W_0W_1-W'_1\right)+\alpha ^2
\left(2W_0W_2+W_1^2-W'_2\right)\nonumber\\
&+& \alpha ^3 \left(2W_0W_3+2W_1W_2-W'_3\right)+\alpha ^4
\left(2W_0W_4+2W_1W_3+W_2^2-W'_4\right)+\ldots .\label{V-W1}
\end{eqnarray}
Secondly, the original eigenvalue $E$ must also be written as
\begin{eqnarray}\sum_{n=0}^{\infty}E_{0,n;m}\alpha^n\end{eqnarray}
where there are three lower indices in the parameter $E_{0,n;m}$
with the index $0$ refereing to the original ground state energy,
and the other index $n$ meaning its nth term expanded in the series
in parameter $\alpha$ and the last $m$ indicating the parameter
$-m^2$ in Eq.(\ref{main Eq2}). One can write the perturbation
equation as
\begin{eqnarray}
W^2-W'&=&V(z,\alpha, m)+m^2=-\left[E{\rm sech}^2z+ \alpha{\rm sech}^2z-\alpha{\rm sech} ^4z\right]+m^2\\
&=&-\left[\sum_{n=0}^{\infty}\alpha^nE_{0n}{\rm sech}^2z+\alpha{\rm
sech}^2z- \alpha{\rm sech} ^4z\right]+m^2. \label{2potential m=s=0
alpha z} \end{eqnarray} Comparing Equations (\ref{potential and w
relation in z}), (\ref{V-W1}), and (\ref{2potential m=s=0 alpha z}),
one could get
\begin{eqnarray}
W_0^2-W'_0&=&-E_{0,0;m}{\rm sech}^2z+m^2\label{0-term}\\
2W_0W_1-W'_1&=& {\rm sech}^4z -(E_{0,1;m}+1){\rm sech}^2z\label{1-term}\\
2W_0W_2+W_1^2-W'_2&=&-E_{0,2;m}{\rm sech}^2z\label{2-term}\\
2W_0W_3+2W_1W_2-W'_3&=&-E_{0,3;m}{\rm sech}^2z\label{3-term}\\
&&\ \ \ \ \ \ \ \ \ldots\nonumber \\
2W_0W_n+\sum_{k=1}^{n-1}W_kW_{n-k}-W'_n&=&-E_{0,n;m}{\rm
sech}^2z\label{4-term}
\end{eqnarray}
From Eq.(\ref{0-term}), we get
\begin{equation}
W_0=m\tanh z,\ E_{0,0;m}=m(m+1).\label{w-0}
\end{equation}
Then, we can write the other equations more concisely
\begin{eqnarray}
W'_1-2m\tanh z W_1&=&(E_{0,1;m}+1){\rm sech}^2z-{\rm sech}^4z\label{1-1-term}\\
W'_2-2m\tanh z W_2&=&E_{0,2;m}{\rm sech}^2z+W_1^2\label{2-1-term}\\
W'_3-2m\tanh z W_3&=&E_{0,3;m}{\rm sech}^2z+2W_1W_2\label{3-1-term}\\
\ \ \ \ \ \ \ \ \ldots\nonumber\\
W'_n-2m\tanh z W_n&=&E_{0,n;m}+\sum_{k=1}^{n-1}W_kW_{n-k}\label{k-1-term}\\
\end{eqnarray}
After obtaining the zero term $W_0$ for the super-potential $W$,
the first order $W_1$ can be gotten as
\begin{equation}
W_1=\bar{A}_1\cosh^{2m}z
\end{equation}
with\begin{eqnarray}\frac{d \bar{A}_1}{d \theta}&=& {\rm sech}^{2m}z \left[(E_{0,1;m}+1){\rm sech}^2z-{\rm sech}^4z\right]\\
\bar{A}_1&=&\int \left[(E_{0,1;m}+1){\rm sech}^{2m+2}z-{\rm
sech}^{2m+4}z\right]dz \end{eqnarray} In order to simplify the
calculation, we just write some useful formula\cite{grad}
\begin{eqnarray}Q(m,z)&=&\int {\rm sech}^{2m}zdz\nonumber\\&=&
\frac{\sinh z}{2m-1}\left[{\rm
sech}^{2m-1}z+\sum_{k=1}^{m-1}\frac{2^k(m-1)(m-2)\ldots(m-k){\rm
sech}^{2m-2k-1}}{(2m-3)(2m-5)\ldots(2m-2k-1)}\right]\end{eqnarray}
This formula could be written in concise form
\begin{eqnarray}Q(m,z)= \frac{\sinh z }{m}\sum_{k=0}^{m-1}I(m,k){\rm
sech}^{2m-2k-1}z,\label{q-m relations}\end{eqnarray} where
\begin{eqnarray}
I(m,k)=\frac{2^km(m-1)(m-2)\ldots(m-k)}{(2m-1)(2m-3)(2m-5)\ldots(2m-2k-1)}.\end{eqnarray}
Hence, \begin{eqnarray}Q(m+n,z)&=& \int {\rm
sech}^{2(m+n)z}dz\nonumber\\ &=& \frac{\sinh z
}{m+n}\sum_{k=0}^{m+n-1}I(m+n,k){\rm sech}^{2(m+n)-2k-1}z\\
&=&Q_1(m+n,z)+2(m+1)I(m+n,n-2)Q(m+1,z)\label{useful
Q-nm,m1}\end{eqnarray} where \begin{eqnarray}Q_1(m+n,z)=\frac{\sinh
z }{m+n}\sum_{j=1}^{n-1}I(m+n,n-1-j){\rm
sech}^{2m+2j+1}z\label{useful Q-nm,Q1}\end{eqnarray} See the
appendix 1 for details.
Hence, the quantity $\bar{A}_1$ is \begin{eqnarray}\bar{A}_1&=&(E_{0,1;m}+1)Q(m+1,z)-Q(m+2,z)\nonumber\\
&=&
\left(E_{0,1;m}+1-2(m+1)I(m+2,0)\right)Q(m+1,z)-Q_1(m+2,z)\nonumber\\
&=&a_1Q(m+1,z)-Q_1(m+2,z)\end{eqnarray} where the introducing
$a_1=E_{0,1;m}+1-2(m+1)I(m+2,0)$ is used. So\begin{eqnarray}
W_1&=& \bar{A}_1\cosh^{2m}z\nonumber\\
&=& a_1Q(m+1,z)\cosh^{2m}z-Q_1(m+2,z)\cosh^{2m}z\end{eqnarray}
 The quantity $E_{0,1;m}$ needs to be determined by the boundary
 conditions that the $\Theta$ is finite at $\pm \infty$.
 This in turn require to calculate the ground eigenfunction upon to the first order
  by
\begin{eqnarray}\Theta_0&=&N\exp\left[-\int Wdz\right]\\&=& N\exp\left[-\int
W_0dz\ -\alpha \int
 W_1dz\right]*\exp{O(\alpha^2)}\\
&=& N{\rm sech}^{m}z \exp\left[-\alpha \int
 W_1dz\right]*\exp{O(\alpha^2)}\label{f-w relation}. \end{eqnarray}
Whenever the eigenfunction is obtained, the boundary conditions
finite $\Psi|_{\pm \infty}$  would choose the proper $E_{0,1;m}$.
 it is easy to compute
\begin{eqnarray}\int W_1dz=\int a_1Q(m+1,z)\cosh^{2m}z dz-\int
Q_1(m+2,z)\cosh^{2m}z dz\end{eqnarray} By the relations (\ref{q-m
relations}),(\ref{useful Q-nm,m1}),(\ref{useful Q-nm,Q1}) it is easy
to obtain the eigenfunction
\begin{eqnarray}\Theta_0&=&N{\rm sech}^{m}z \exp\left[\alpha
\left(a_1\sum_{k=0}^{m}\frac{I(m+1,k)\cosh^{2k}z }{2k(m+1)}\right)\right]\nonumber\\
&& *\exp\left[ \alpha
 \frac{I(m+2,0)\rm{sech}^{2}z}{2m+4}\right]*\exp{O(\alpha^2)}\end{eqnarray}
Unless the coefficient $a_1$ of the term
$\sum_{k=0}^{m}\frac{I(m+1,k)\cosh^{2k}z }{2k(m+1)}$ becomes zero,
the eigenfunction will blow up at infinity whenever $a_1\alpha<0,$.
So, $E_{0,1;m}$ is determined by $a_1=E_{0,1;m}+1-2(m+1)I(m+n,0)=0$,
the results are
\begin{eqnarray}
E_{0,1;m}&=&-\frac1{2m+3}\label{e1 in simple }\\
 W_1&=&-Q_1(m+2,z)\cosh^{2m}z=-\frac{\sinh z }{m+2}I(m+2,0){\rm
sech}^{3}z\nonumber\\ &&=-\frac{\sinh z {\rm
sech}^{3}z}{2m+3}\label{good result of w1}\\
 \Theta_0&=&N{\rm sech}^{m}z\exp\left[ \alpha
 \frac{\rm{sech}^{2}z}{4m+6}\right]*\exp{O(\alpha^2)}.\end{eqnarray}

With the first order term of the super-potential $W_1$, one could
compute the second term $W_2$ by the same process. \begin{eqnarray}
W_2=\bar{A}_2\cosh^{2m}z\end{eqnarray} with
\begin{eqnarray}\bar{A}_2&=&\int
{\rm sech}^{2m}z\left(E_{0,1;m}{\rm sech}^2z+W_1^2\right)dz\nonumber\\
&=&E_{0,1;m}Q(m+1,z)+\int {\rm sech}^{2m}z \frac{\sinh^2 z
{\rm sech}^{6}z}{(2m+3)^2}dz\nonumber\\&=&E_{0,1;m}Q(m+1,z)+\frac{1}{(2m+3)^2}\left(Q(m+2,z)-Q(m+3,z)\right)\nonumber\\
&=&\left[E_{0,1;m}+\frac{2(m+1)}{(2m+3)^2}\left(I(m+2,0)-I(m+3,1)\right)\right]Q(m+1,z)\nonumber\\
&&
+\frac{1}{(2m+3)^2}\left[Q_1(m+2,z)-Q_1(m+3,z)\right]\end{eqnarray}
Similarly, the coefficient
$a_2=E_{0,1;m}+\frac{2(m+1)}{(2m+3)^2}\left(I(m+2,0)-I(m+3,1)\right)$
of the term $Q(m+1,z)$ must zero. The results are elegant:
\begin{eqnarray}
E_{0,2;m}&=&-\frac{2m+2}{(2m+3)^3(2m+5)}\label{e2 in simple }\\
 W_2&=&
\left[\frac{\sinh z{\rm sech}^3z}{(2m+3)^3(2m+5)}-\frac{\sinh z {\rm
sech}^5z}{(2m+3)^2(2m+5)}\right] \label{w2 in complex
1}.\end{eqnarray}

We can continue to calculate $W_3,\ W_4,\ \ldots,\ W_n,\ \ldots$.
here we just use induction to prove that \begin{eqnarray}W_n=\sinh
z\sum_{k=1}^{n}a_{n,k}{\rm sech}^{2k+1}z\label{w-form}\end{eqnarray}
with $a_{n,k}$ could be determined by $a_{i,j},i<n,j<n$, which
appear in $W_i,\ i<n$. Obviously, $W_1$ in Eq.(\ref{good result of
w1}) satisfies the requirement (\ref{w-form}) in the case $n=1$.
Suppose that $W_k,\ k<n$ meets the requirement, then
\begin{eqnarray}W_{n}=\bar{A}_{n}\cosh^{2m}\end{eqnarray} with $\bar{A_n}$  is determined by
\begin{eqnarray}&& \bar{A}_{n}=\int
{\rm sech}^{2m}z\left[E_{0,n;m}{\rm
sech}^2z+\sum_{k=1}^{n-1}W_kW_{n-k}\right]dz.\end{eqnarray} The
first thing is to simplify the following term:
\begin{eqnarray}
\sum_{k=1}^{n-1}W_kW_{n-k}&=&\sum_{k=1}^{n-1}\sum_{i=1}^{k}\sum_{j=1}^{n-k}
a_{k,i}a_{n-k,j}\sinh^2z{\rm sech}^{2(i+j)+2}z\nonumber\\
&=& \sum_{k=1}^{n-1}\sum_{p=2}^{n}\sum_{j=1}^{p-1}
a_{k,p-j}a_{n-k,j}\sinh^2z{\rm sech}^{2p+2}z\nonumber\\
&=&\sum_{k=1}^{n-1}\sum_{p=2}^{n}\sum_{j=1}^{p-1}
a_{k,p-j}a_{n-k,j}\left[{\rm sech}^{2p}z-{\rm
sech}^{2p+2}z\right]\nonumber\\&=&\sum_{k=1}^{n-1}\sum_{p=3}^{n}\sum_{j=1}^{p-1}
\left[a_{k,p-j}a_{n-k,j}-a_{k,p-1-j}a_{n-k,j}\right]{\rm sech}^{2p}z
\nonumber\\ && +\sum_{k=1}^{n-1}\sum_{j=1}^{2}
a_{k,2-j}a_{n-k,j}{\rm sech}^{4}z-\sum_{k=1}^{n-1}\sum_{j=1}^{n}a_{k,n-j}a_{n-k,j}{\rm sech}^{2n+2}z\nonumber\\
&=&\sum_{p=2}^{n+1}b_{n,p}{\rm sech}^{2p}z\\
b_{n,p}&=&\sum_{k=1}^{n-1}\sum_{j=1}^{p-1}
\left[a_{k,p-j}a_{n-k,j}-a_{k,p-1-j}a_{n-k,j}\right],p=2,3,\ldots,n+1\end{eqnarray}
with the conditions $i<j,\ or j<1 $ then \begin{eqnarray}a_{i,j}=0
\end{eqnarray}

\begin{eqnarray} \bar{A}_{n}&=&E_{0,n;m}Q(m+1,z)+\sum_{p=2}^{n+1}b_{n,p}Q(m+p,z)\nonumber\\
&=&\left[E_{0,n;m}+2\sum_{p=2}^{n+1}b_{n,p}\frac{m+1}{m+p}I(m+p,p-2)\right]Q(m+1,z)\nonumber\\&&+\sum_{p=2}^{n+1}b_{n,p}Q_1(m+p,z)\end{eqnarray}
In order to make the eigenfunction finite at infinity, the quantity
$E_{0,m,n+1}$ is selected by \begin{eqnarray}&&
E_{0,n;m}+2\sum_{p=2}^{n+1}b_{n,p}(m+1)I(m+p,p-2)=0,\nonumber\end{eqnarray}
that is,
\begin{eqnarray}E_{0,n;m}=-2\sum_{p=2}^{n+1}b_{n,p}(m+1)I(m+p,p-2).\label{e0nm}\end{eqnarray}
hence, \begin{eqnarray}W_{n}&=&\bar{A}_n\cosh^{2m} z=\sum_{p=2}^{n+1}b_{n,p}Q_1(m+p,z)\cosh^{2m}z\nonumber\\
&=& \sum_{p=2}^{n+1}b_{n,p}\frac{\sinh z}{m+p}
\sum_{j=1}^{p-1}I(m+p,p-1-j){\rm sech}^{2j+1}z\nonumber\\
&=& \sinh z
\sum_{j=1}^{n}\sum_{p=j+1}^{n+1}b_{n,p}\frac{I(m+p,p-1-j)}{m+p}{\rm
sech}^{2j+1}z \nonumber\\ &=&\sinh z\sum_{j=1}^{n}a_{n,j}{\rm
sech}^{2j+1}z \end{eqnarray}

\begin{eqnarray}a_{n,j}=\sum_{p=j+1}^{n+1}b_{n,p}\frac{I(m+p,p-1-j)}{m+p}\end{eqnarray} we see
that $a_{n,j}$ are determined by $b_{n,p}$, which are completely
defined by $a_{i,j},i<n,j<n$. this completes our proof. Therefore,
the super-potential $W$ could be written as \begin{eqnarray}
W&=&W_0+\sum_{n=1}^{\infty}W_n\alpha^n=W_0+\sum_{n=1}^{\infty}\sinh
z\sum_{j=1}^{n}a_{n,j}{\rm sech}^{2j+1}z \alpha^n\label{all wn in
z}\end{eqnarray}

 The ground eigenfunction becomes \begin{eqnarray}
\Theta_0&=&N\exp\left[-\int Wdz\right]\nonumber\\&=&
N\exp\left[-\int W_0dz\ -\sum_{n=1}^{\infty}\alpha^n \int
 W_ndz\right]\nonumber\\&=& N\exp\left[-\int m\tanh zdz
-\int \sum_{n=1}^{\infty}\sum_{j=1}^{n}\alpha^n a_{n,j}\sinh z{\rm
sech}^{2j+1}z dz\right]\nonumber\\&=&N{\rm
sech}^{m}z\exp\left[\sum_{n=1}^{\infty}\sum_{j=1}^{n}\frac{\alpha^n
a_{n,j}}{2j}{\rm sech}^{2j}z \right].\end{eqnarray} Back to the
independent variable $\theta$, ${\rm sech} z=\sin\theta$, the above
ground eigenfunction becomes \begin{eqnarray}
\Theta_0=N\sin^{m}\theta\exp\left[-\sum_{n=1}^{\infty}\sum_{j=1}^{n}\frac{\alpha^n
a_{n,j}}{2j}\sin^{2j}\theta \right]\label{the same result of the
ground function }\end{eqnarray} The original ground eigen-energy is
\begin{eqnarray}
E_{0;m}=m(m+1)+\sum_{n=1}^{\infty}\alpha^nE_{0,n;m}\label{e0m}\end{eqnarray}
with $E_{0,n;m}$ is determined by Eq.(\ref{e0nm}).
\section*{Comparison with the former results }
The elegant forms (\ref{the same result of the ground function
}),(\ref{e0m}),(\ref{e0nm})for the ground eigenfunction and
eigenvalue are the same as that in the reference \cite{tian
general}. There appeared two forms for the super-potential: $W(z)$
of the form (\ref{all wn in z}) and $W(\theta)$ of (\ref{wn to be
proved}) in the reference \cite{tian general}. Whatever forms may
the super-potentials be, they should give the same eigenfunctions
for the spheroidal equations. Thus, it results the relation between
them as
\begin{eqnarray}\Theta_0(\theta)=\int W(z)\frac{dz}{d\theta}d\theta.\end{eqnarray} By
$z=\lg\tan\frac{\theta}2$, it is easy to get
\begin{eqnarray}W(\theta)=W(z)\frac{dz}{d\theta}=\frac1{\sin\theta}W(z).\end{eqnarray} Writing
$W(z)$ back as the function of the original independent variable
$\theta$, one gets
\begin{eqnarray}W(\theta)&=&W_0(\theta)+\sum_{n=1}^{\infty}W_n(\theta)\alpha^n\\
W_0(\theta)&=&-m \cot\theta, \\
W_n(\theta)&=&W_n=-\cos\theta
\sum_{k=1}^{n}a_{n,k}\sin^{2k-1}\theta\end{eqnarray} Comparing the
results with that (\ref{wn to be proved}) of the results of the
reference \cite{tian general}, they are the same except for the
$W_0(\theta)$. The difference between them is originated from the
eigenfunction's relation \begin{eqnarray}
\Theta_0 &=& \frac{\Psi_0}{\sin^{\frac12}\theta}\\
\Psi_0 &=& \int W(\theta)\theta.\end{eqnarray} where $W(\theta)$ are
the forms of (\ref{wn to be proved}). The advantage of the method is
that it can be easily extended to the spin-weighted spheroidal
equations. It will be our further study.
\section*{Obviously non-shape-invariance property of the spheroidal functions}
In order to check whether the spheroidal wave equations have the
shape-invariance property, the super-potential $W$ is rewritten as
the following form: \begin{eqnarray} W(A_{n,j},z)=A_{0,0}m\tanh z
+\sum_{n=1}^{\infty}\sinh z\sum_{j=1}^{n}A_{n,j}a_{n,j}{\rm
sech}^{2j+1}z \alpha^n,\end{eqnarray} With the definition
\begin{eqnarray} W_n(A_{i,j},z)=\sinh
z\sum_{j=1}^{n}A_{n,j}a_{n,j}{\rm sech}^{2j+1}z
\end{eqnarray} Then, $V^{\pm}(A_{n,j})$ are defined as
\begin{eqnarray} V^{\pm}(A_{n,j},z)=W^2(A_{n,j},z)\mp W'
=\sum_{n=0}^{\infty}\alpha^nV^{\pm}_n(A_{i,j},z).\end{eqnarray} We
will check whether or not $V^{\pm}_n(A_{i,j},z)$ are related with by
the relations \begin{eqnarray}
V^{+}_n(A_{i,j},z)=V^{-}_n(B_{i,j},z)\end{eqnarray} step by step.

First, we write the special cases for $V_n^{\pm}, \ n=0,\ 1$. When
$n=0$,\begin{eqnarray}
&&V_0^-=W_0^2(A_{0,0},z)-W_0^2(A_{0,0},z)=m^2A_{0,0}^2\tanh^2z-A_{0,0}m{\rm sech}^2z\\
&&=m^2A_{0,0}^2-(m^2A_{0,0}^2+A_{0,0}m){\rm sech}^2z\\
&&V_0^+=W_0^2(A_{0,0},z)+W_0^2(A_{0,0},z)=m^2A_{0,0}^2\tanh^2z+A_{0,0}m{\rm sech}^2z\\
&&=m^2A_{0,0}^2-(m^2A_{0,0}^2-A_{0,0}m){\rm sech}^2z\\
&&=V_0^-(B_{0,0},z)=m^2B_{0,0}^2-(m^2B_{0,0}^2+B_{0,0}m){\rm sech}^2z+R_0(A_{0,0})\\
&& R_0(A_{0,0})=m^2A_{0,0}^2-m^2B_{0,0}^2\end{eqnarray} Therefore
\begin{eqnarray} B_{0,0}=A_{0,0}-\frac1m,\
R_0(A_{0,0})=2mA_{0,0}-1\label{shape-invariance 0}.\end{eqnarray}
This results is exact when $\alpha=0$ , and it just shows that the
associated Legendre equations have the shape-invariance properties.
It is easy to get the recurrence relations (\ref{p-n m fixed}) for
the associated Legendre functions from the the shape-invariance
properties. What interests us most is whether this property could
extend to the spheroidal functions, the case $\alpha \ne 0$.

 When
$n=1$,
\begin{eqnarray} V_1^-(A_{0,0},A_{1,1},z)&=&2W_0(A_{0,0})W_1(A_{1,1},z)-W'_1(A_{1,1},z)\nonumber\\
&=&-\frac{2mA_{0,0}A_{1,1}}{2m+3}\sinh^2z{\rm
sech}^4z+\frac{A_{1,1}}{2m+3}\left[\sinh
z{\rm sech}^3z\right]'\nonumber\\
&=&\frac{(2mA_{0,0}+3)A_{1,1}}{2m+3}{\rm sech}^4z+\frac{(2-2mA_{0,0})A_{1,1}}{2m+3}{\rm sech}^2z\\
V_1^+(A_{0,0},A_{1,1},z)&=&2W_0(A_{0,0})W_1(A_{1,1},z)+W'_1(A_{1,1},z)\nonumber\\
&=&-\frac{2mA_{0,0}A_{1,1}}{2m+3}\sinh^2z{\rm
sech}^4z+\frac{A_{1,1}}{2m+3}\left[\sinh
z{\rm sech}^3z\right]'\nonumber\\
&=&\frac{(2mA_{0,0}-3)A_{1,1}}{2m+3}{\rm
sech}^4z+\frac{(-2-2mA_{0,0})A_{1,1}}{2m+3}{\rm
sech}^2z\end{eqnarray} If we require the first term with the
shape-invariance property, that is, \begin{eqnarray}
V_1^+(A_{0,0},A_{1,1},z) &=&V_1^-(B_{0,0},B_{1,1},z)\nonumber\\&=&
\frac{(2mB_{0,0}+3)B_{1,1}}{2m+3}{\rm
sech}^4z+\frac{(2-2mB_{0,0})B_{1,1}}{2m+3}{\rm
sech}^2z,\end{eqnarray} there exist two conditions for the quantity
$B_{1,1}$ to meet:
\begin{eqnarray}
&&(2mB_{0,0}+3)B_{1,1}=(2mA_{0,0}-3)A_{1,1}\label{useful b11}\\
&& (2-2mB_{0,0})B_{1,1}=(-2-2mA_{0,0})A_{1,1}\label{useless
b11}.\end{eqnarray} It is easy to see that these two conditions are
not compatible, one is led to the conclusion the shape-invariance
property is not hold for the spheroidal functions. first,  we dot
not admit them and investigate carefully. The shape-invariance
property of the zero-term (\ref{shape-invariance 0}) shows the
spheroidal functions have the same property as that of the
associated Legendre functions. The new eigenvalue of the
Eq.(\ref{main Eq2}) is the quantity $-m^2$, whereas the original
eigenvalue $E$ now is contained in the expression $E{\rm sech}^2z$.
Therefore, when the new eigenvalue changes from $-m^2$ to
$-(m-1)^2$, the original eigenvalue $E$ can not remains unchanged.
this is the great difference of the spheroidal functions from the
the associated Legendre functions where the original eigenvalue
remains the same as the new eigenvalue changes from $-m^2$ to
$-(m-1)^2$. So the potential have no shape-invariance property.

\section*{Acknowledgements}
This work was supported in part by the National Science Foundation
of China under grants No.10875018, No.10773002.
\section*{Appendix1: Simplification of the quantity $Q(m+n,z)$}

\ \ \ \ \begin{eqnarray}Q(m+n,z)&=& \int {\rm
sech}^{2(m+n)z}dz=\frac{\sinh z
}{m+n}\sum_{k=0}^{m+n-1}I(m+n,k){\rm sech}^{2(m+n)-2k-1}z\nonumber\\
&=& Q_1(m+n,z)+Q_2(m+n,z)\end{eqnarray} where the two parts
$Q_1(m+n,z),Q_2(m+n,z)$ are:\begin{eqnarray}Q_1(m+n,z)&=&\frac{\sinh
z }{m+n}\sum_{k=0}^{n-2}I(m+n,k){\rm
sech}^{2(m+n)-2k-1}z\\Q_2(m+n,z)&=& \frac{\sinh
z}{m+n}\sum_{k=n-1}^{m+n-1}I(m+n,k){\rm
sech}^{2(m+n)-2k-1}z.\end{eqnarray} In the above equation for the
quantity $Q_1$,  the change  of $j=n-1-k$ simplifies it as
\begin{eqnarray}Q_1(m+n,z)&=&\frac{\sinh z
}{m+n}\sum_{j=1}^{n-1}I(m+n,n-1-j){\rm
sech}^{2m+2j+1}z.\label{Q1}\end{eqnarray} The second part
$Q_2(m+n,z)$is connected with $Q(m,z)$. Here is the proof. First
changing $k$ in the above equation for the quantity $Q_2$ to
$p=k-(n-1)$, it is easy to obtain
\begin{eqnarray}Q_2(m+n,z)&=& \frac{\sinh
z}{m+n}\sum_{p=0}^{m}I(m+n,p+(n-1)){\rm sech}^{2m-2p+1}z\nonumber\\
&=&\frac{\sinh z}{m+n}\sum_{p=0}^{m}I(m+1+(n-1),p+(n-1)){\rm
sech}^{2m-2p+1}z\end{eqnarray} here the relations between
$I(m+l,k+l)$ and $I(m,k)$ are \begin{eqnarray}
&&I(m+l,k+l)\nonumber\\
&=&\frac{2^{k+l}(m+l)(m+l-1)\ldots
m(m-1)(m-2)\ldots(m-k)}{(2m+2l-1)(2m+2l-3)\ldots(2m-1)(2m-3)(2m-5)\ldots(2m-2k-1)}\nonumber\\
&=& \frac{2*2^{l-1}(m+l)(m+l-1)\ldots
(m+1)}{(2m+2l-1)(2m+2l-3)\ldots(2m+1)}\frac{2^k
m(m-1)(m-2)\ldots(m-k)}{(2m-1)(2m-3)\ldots(2m-2k-1)}\nonumber\\&=&
2I(m+l,l-1)I(m,k).\end{eqnarray} So, \begin{eqnarray}
I(m+1+(n-1),p+(n-1))=2I(m+1+(n-1),n-2)I(m+1,p)\end{eqnarray}

The quantity $Q_2(m+n,z)$ becomes
\begin{eqnarray}Q_2(m+n,z)&=&\frac{\sinh
z}{m+n}\sum_{p=0}^{m}2I(m+1+(n-1),(n-2))I(m+1,p){\rm
sech}^{2m-2p+1}z\\&=&2I(m+1+(n-1),(n-2)) \frac{\sinh
z}{m+n}\sum_{p=0}^{m}I(m+1,p){\rm sech}^{2m-2p+1}z \nonumber\\&=&
2I(m+n,n-2)(m+1)Q(m+1,z).\end{eqnarray} Therefore, \begin{eqnarray}
Q(m+n,z)=Q_1(m+n,z)+2(m+1)I(m+n,n-2)Q(m+1,z).\label{useful Q-nm,m1
appendix }\end{eqnarray}

\end{document}